\documentstyle[12pt,aps,epsf,preprint,tighten]{revtex} 

\begin{document} 
\draft 
\title{ 
\begin{flushright}
{\small Preprint HKBU-CNS-9731\\
November 1997}
\end{flushright}
Energy transport and correlation between two attractors 
connected by Fermi-Pasta-Ulam chain 
} 
\author{  
Alexander Fillipov$^{[1,2]}$, Bambi Hu$^{[1,3]}$, Baowen 
Li$^{[1]}$, and  Alexander Zeltser$^{[1,2]}$}  
\address{ $^{[1]}$ Department of Physics and Center for Nonlinear 
Studies, Hong Kong  Baptist University, Hong Kong, China \\ 
$^{[2]}$ Donetsk Physical-Technical Institute, National Academy of  
Science, 340114, Donetsk,Ukraine\\ $^{[3]}$ Department of Physics, 
University of Houston, Houston TX 77204} 
\date{\today}  
\maketitle 
 
\begin{abstract} 
We study numerically time evolution of a system which consists of 
two  attractors connected by Fermi-Pasta-Ulam (FPU) chain. It is 
found that after sufficiently long time there exits 
self-consistent large 
scale structure in the system.
The wavelet transform is used to 
separate the modes in different scales.  We found that the
nonlinear long wavelength mode propagating along the chain assists  
the energy transport. Further, all points in the system are 
found to be correlated. Our results 
explain satisfactorily why the chaotic behaviour is not 
enough to ensure 
the Fourier heat law and the thermal conductivity diverges as it is found 
recently by Lepri {\em et al} (PRL {\bf 78}, 1897(1997).).  
 
\end{abstract}  

The  finding of 
Fermi-Pasta-Ulam\cite{FPU55} in 1955, i.e. the absence of the energy 
equipartition in a system of coupled 
nonlinear oscillators, becomes one of the cornerstones in the modern 
statistical mechanics \cite{Ford92,LL92}. It stimulated the study of 
nonlinear dynamics and chaos.

Since the first numerical experiment of FPU , many works have been done 
on one-dimensional anharmonic oscillators to study various problems related 
to irreversiblie statistical mechanics \cite{Ford92}. 
In addition to many applications of  FPU model to study the relation 
between the stochastic motions and thermodynamics properties 
\cite{Kantz,Pettini}, the FPU model is used very recently to study 
another important problem in nonequilibrium system by Lepri 
{\em et al}\cite{Lepri97} 
(thereafter refereed to as LLP), namely the Fourier heat conduction 
law in insulating solids. In LLP's study, two Nos$\acute{e}$-Hoover 
"thermostats" \cite{NH841,NH842}  were put on the first and 
last particles of FPU-$\beta$ model (FPU model with quartic 
potential) keeping the temperature at $T_+$ and $T_-$, respectively. 
The two ends of the chain are fixed. After a certain transient time, the 
nonequilibrium stationary state sets in and a nonlinear shape of 
temperature profile is formed. The temperature is found to have 
scaling relation $T_l=T(l/N)$, here $N$ is the number of the particles. 
The heat flux is found to be $N^{1/2}$. Therefore, the thermal 
conductivity $\kappa$ is divergent approximately with the length of 
the chain as $N^{1/2}$, and the Fourier heat law is not 
justified. Compared with that case of the ding-a-ling model 
\cite{Casati84}, where the heat transport obeys the  Fourier  heat law, 
LLP concluded that the chaotic behaviour is not sufficient to ensure the 
Fourier heat law. 
 
Although it is commonly believed that the anharmonicity or the 
nonlinear interaction leads to the scattering of phonons, the mechanism 
leading to the 
divergence of the thermal conductivity in FPU model and other similar 
nonlinear oscillator chains is still lacking.  In fact, it is a very 
general problem in the field of statistical physics to understand the 
origin of the irreversibility and its compatibility with the time 
reversible deterministic microscopic dynamics.  
 
In this paper, concentrating on time evolution of system used by LLP, 
we would  like to clarify the mechanism behind the divergence of the 
thermal conductivity. The equations of motion of the two particles 
keeping at the "thermostats" are determined by, 
\begin{equation} 
\begin{array}{l} 
\ddot{x}_1 = -\zeta_+ \dot{x}_1 + f_1 - f_2,\\ 
\ddot{x}_N = -\zeta_- \dot{x}_N + f_N - f_{N+1},\\ 
\dot{\zeta}_+ = \frac{\dot{x}^2_1}{T_+} -1,\qquad 
\dot{\zeta}_- = \frac{\dot{x}^2_N}{T_-} -1. 
\end{array} 
\label{eqm1} 
\end{equation} 
The equation of motion for the central particles is, 
\begin{equation} 
\ddot{x}_i = f_i - f_{i+1},\qquad i=2,\dots,N-1, 
\label{eqm2} 
\end{equation} 
where $f_i = -V'(x_{i-1}-x_i)$ is the force acting on the particle, 
$V(x) = x^2/2+\beta x^4/4$. $x_0=0$ and $x_{N+1} =0$. It is 
obvious, the dynamical equations are invariant under time reversal 
combined with the change $p_i \to -p_i$.  
 
As an alternative, the coupling  of a chain of N particles to thermal 
reservoirs can be realized by numerical random generators 
\cite{HKMM93}. In this case at each time step, the velocities of the 
first and last particles are determined by sampling randomly from the 
Maxwellian distribution corresponding to reservoirs temperatures. 
This approach allows one to achieve quickly steady state for 
time-averaged temperature profile for sufficiently long chain to study 
dependence of thermal conductivity as a function of the system size $N$. 
In its turn, Nos$\acute{e}$-Hoover "thermostats" were proposed 
specially as the extension of molecular dynamics methods to treat the 
problems relating to the question\cite{Ford92,NH841}: 
can statistical mechanics be derived from the underlying dynamics?  
 
It has been shown that the canonical distribution with given 
temperature can be generated with smooth, deterministic and time 
reversible trajectories. However, simple numerical examples given in 
\cite{NH842} illustrate that a single oscillator which is used usually 
for the boundary "thermostat" in the equations (1) can not be 
sufficiently chaotic to yield the canonical distribution from a 
single initial condition. Really, for the parameter we chosen, 
$T_+=152, T_-=24$ (the same as LLP used), and $\beta=0.5$, the time 
evolution of the two ""thermostats" are not sufficiently random. There 
are some periodic-like structures in the phase space (see Fig.1). These two 
"thermostats" are not really thermostats, instead they are  attractors 
covering only part of the phase space. It is not quite clear that even 
large system including these two subsystems as the "boundary 
conditions" will behave quite randomly. 
 
In fact we are dealing with long but finite system driven by two attractors 
which are connected by FPU $\beta$ chain. Energy balance is fixed by 
the two "thermostats" only and dissipation is absent. This means that a 
memory about  some excitations generated once in this system may not 
disappear during thermalization process.  
 
The time evolution of the displacement at initial time 
period ($10^2$ time units) as well as after long enough time (the same 
time period, but after $10^4$ time units) are shown in Fig. 1c,d 
respectively. It is evidently that there are some kind of nonlinear 
excitations 
propagating inside the system, which does not disappear even after long 
time. A combination of the parameters of the chain and "thermostats" 
leads to some mesoscopic structures.  
 
We tested that after sufficiently long averaging this structure is not 
reflected directly on the "temperature" profile. (Temperature is defined as 
the twice of 
the kinetic energy). Namely, the temperature profile is smooth and 
exactly has the scaling $T_l=T(l/N)$ as observed by LLP. However, at least 
two different 
kinds of regular motion alive in the system at sufficiently long time. There 
are quasi-soliton nonlinear excitations performing non-diffusive 
energy transport between two boundary attractors and longwave 
stationary eigen-modes of displacements along the chain.  
 
First ones are nonlinear excitations which move back and forth and 
scattered by the boundaries. Their presence is the main reason for the 
seemly singularity of the temperature profile near the boundary. 
 As an independent test, we initialized the displacement by a 
localized excitation 
and let it propagate along the chain (Fig.2a). We calculate the "temperature" 
profile caused by the propagation of  this excitation (Fig. 2b). It shows 
specific structure near the boundary as displayed in the temperature 
profile in LLP. 
 
The second kind of motion can be compared to that of an effective 
harmonic oscillators with rescaled frequencies and modulated 
amplitudes existing at different time scales. If we take
enough long time for the simulation, then
the oscillations of very long period are observable. 
Besides, the time taken to reach the 
same level of thermalization essentially grows with the chain length $N$. 
Strictly 
speaking, this makes impossible the simulation of the thermodynamic 
limit.
 
Nevertheless, the thermalization-like process is realized in the 
nonequilibrium stationary state. Fig. 3. shows the probability 
distribution function of the velocity of the particles near the boundary 
and that inside the chain far from the boundary. They are really good 
Gaussian function, with the widths proportional to the averaged 
temperatures at the same points.  

Although 
this nonlinear chain is stochastic, it is not stochastic enough to be 
a purely 
random system. Let us plot the time-evolution of the displacement. 
The particular plot in Fig. 4. shows the system having 64 particles. The 
equation of motion have been integrated through a standard fifth-order 
Runge-Kutta routine, using double precision with a maximal step 
equal to $10^{-3}$. The time in Fig 4a begins after a transient 
period of 
$10^8$ steps  and changes from 0 to 163.84 time units. The color 
shows the displacements from most negative ("cold" colors) to most 
positive ("hot") values. Qualitatively the same figures were obtained 
for other chains (from 32 till 1024 particles). It has been checked also 
for  the structures obtained after very long time transient 
periods as 
well as much longer time inteval of observation were used (up to 
1638.4 units).  
 
Strongly pronounced  periodic structure in this picture corresponds to 
the longest oscillation visible during the demonstrating time 
period\footnote{Note that the period of this structure is found to be 
proportional to $N$, whereas that of soliton in FPU model is 
proportional to $N^{5/2}$.}. It shades 
other waves in system. Presence of the excitations with different scale 
makes it impossible to separate them in ordinary Fourier 
transform approach. However, it can be done with help of more 
sophisticated wavelet transform, which allows us to separate the 
excitations in different scales naturally \cite{TA}. To filtrate the 
modes a standard Daubechies wavelets of 20 order DAUB20 have been 
used \cite{NR}. The wavelet transform over time has been performed 
for every particle and truncation is made over first 0.39\% 
wavelet coefficients.  
 
The filtration gives clear picture of the long wavelength motion, shown 
in Fig. 4b. Fig. 4c displays mesoscopic structure which 
remains after subtraction of the above large-scale structure  from the 
complete one shown in Fig 4a. Many short wave excitations 
moving left and right without energy loss are seen clearly. They are 
found to be responsible for instant local heat flux $J_i (t)$ in the 
system. Last value has the interpretation of the flow of potential energy 
from the $i$th to its neighboring particle and can be written in the 
form 
\begin{equation} 
\ J_i(t) =\dot{x}_{i}f_{i+1}. 
\label{eqm3} 
\end{equation} 
The time-space distribution for the $J_i(t)$ value is displayed explicitly 
(without wavelet transform) in Fig. 4d  and should be compared with 
mesoscopic structure in Fig.4c.  
 
The regular energy transport along the chain generates long-time 
correlations between displacements in the system. It is reflected by the 
correlation functions 
\begin{equation} 
\ C_{ij}(t') =\langle{x}_{i}(t) x_{j}(t+t')\rangle 
\label{eqm4} 
\end{equation}  
Fig.5a shows the autocorrelation function $C_{ii}$ (each of them is 
normalized on its maximum).
It is seen from this picture that the $C_{11}$ and $C_{NN}$
have a form qualitatively close to that of independent chaotic 
attractors. The quasi-periodic correlations inside the chain are also 
visible. In addition, an interaction between particles leads to strong 
correlations between main periods of motion for different points, 
including the particles at both ends.   
This is readily depicted by $C_{1N}$ shown in 
Fig. 5b. This plot is quite essential.
It tells us that due to interaction via the chain some regular 
correlation shows up even between the two particles which are 
supposed to be kept at the "thermostats". 
 
In conclusion,  formation of attracting large-scale structure as well as 
the nonlinear 
excitation of propagating waves are the main courses which make 
the energy transport in such a system {\em impossible}
to obey the 
Fourier heat law. The results given here suggest that, in order to have 
energy transport obey the Fourier heat law, we need to add some 
ingredients such as the periodic external potential\cite{HLZ98}, which  
is analogous to the lattice, to 
inhibit such kind of structure and long wavelength propagating modes.  
 
This work was supported in part by the grants from the Hong Kong Research 
Grants  Council (RGC) and the Hong Kong Baptist University Faculty Research  
Grant (FRG).

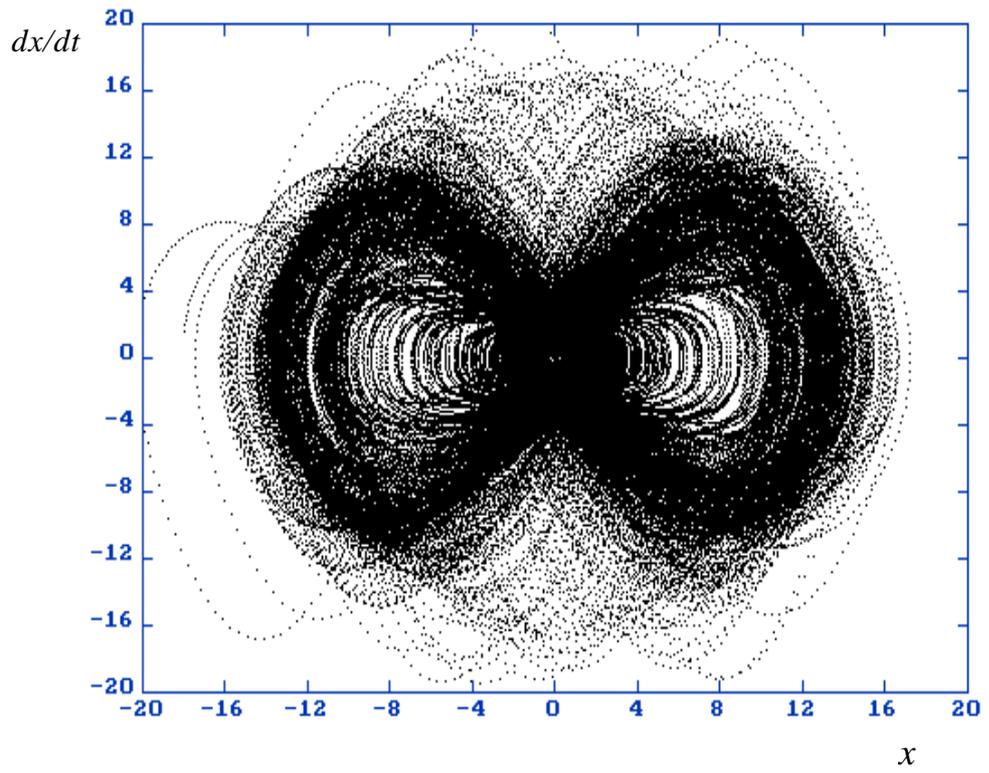
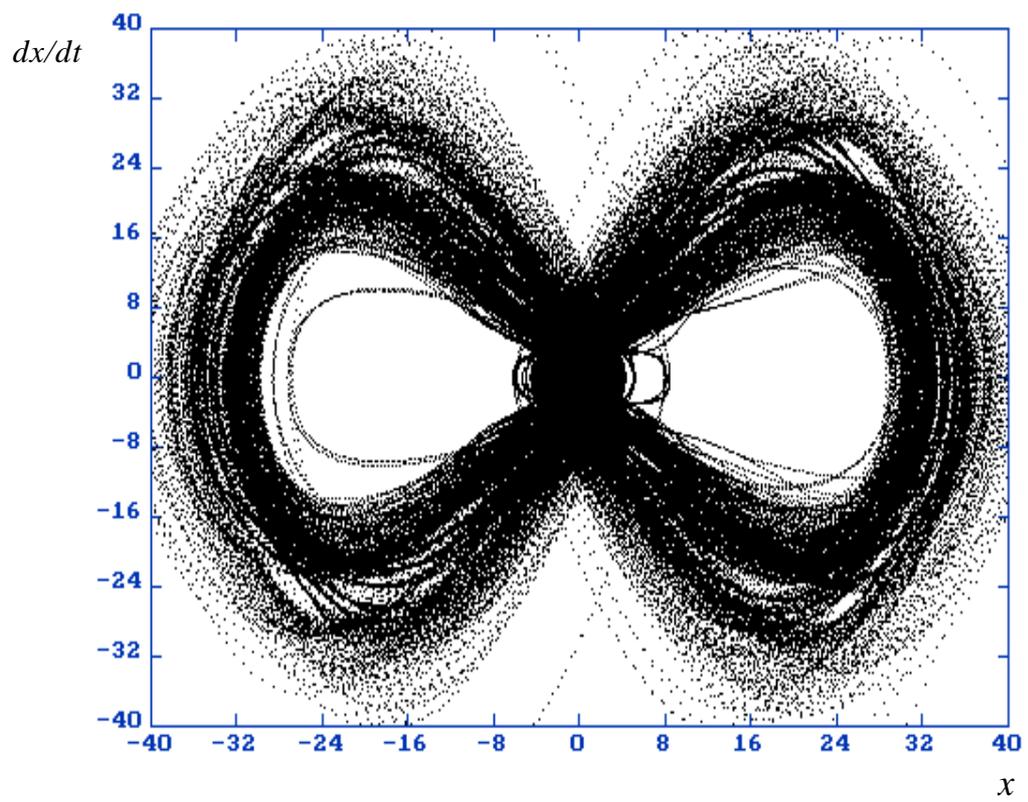
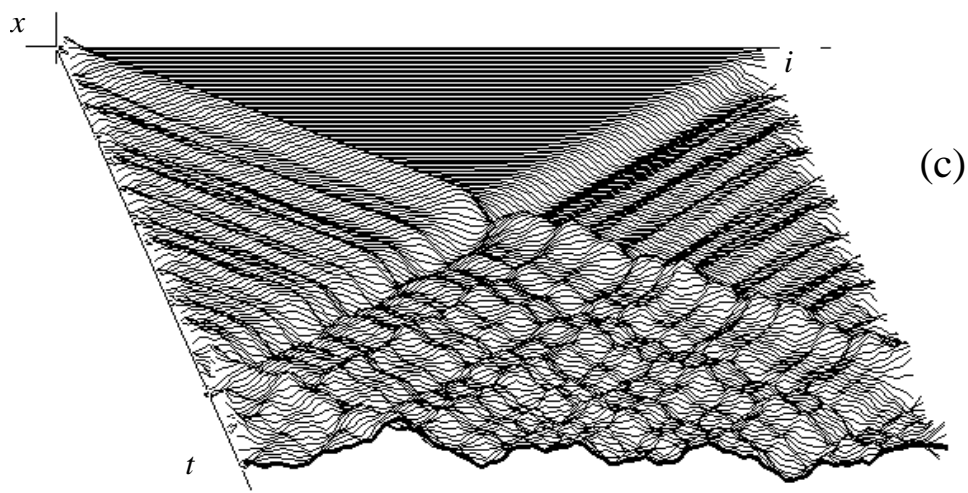
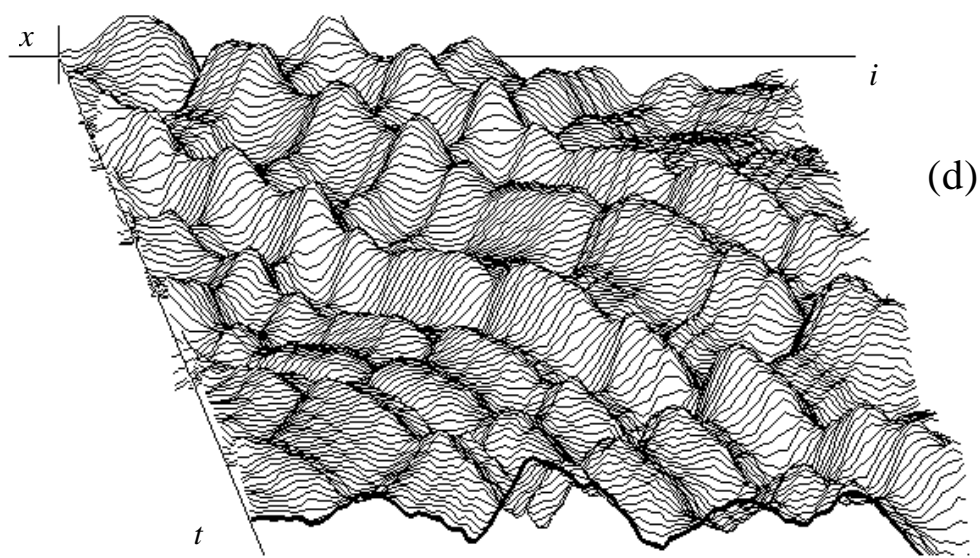
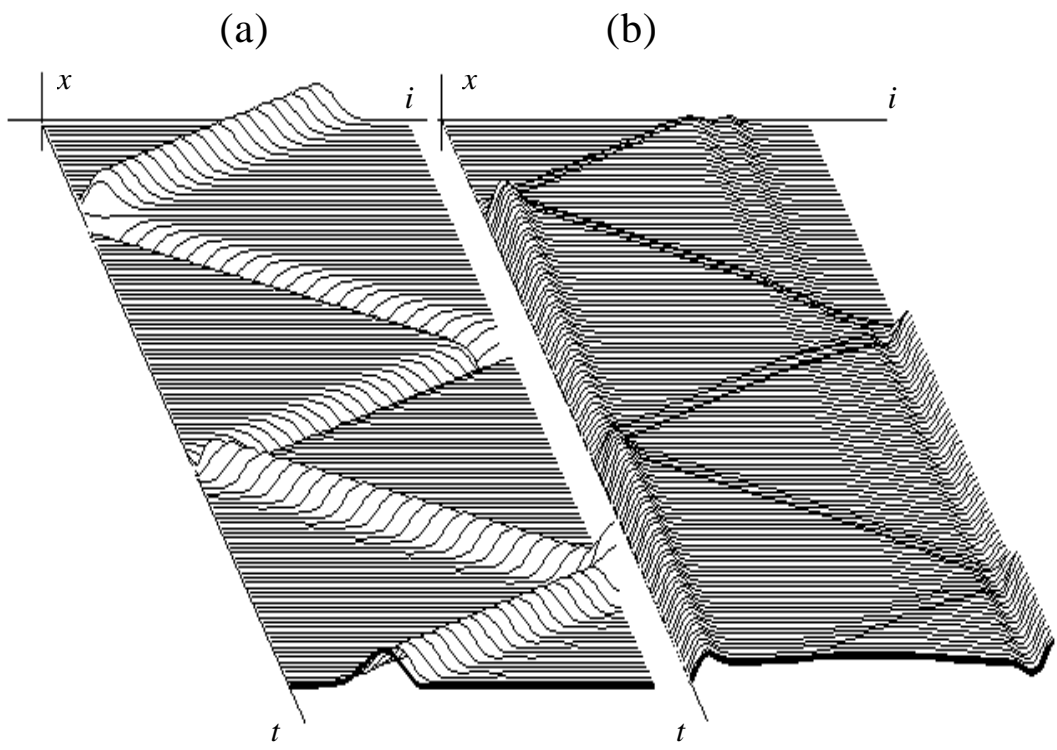
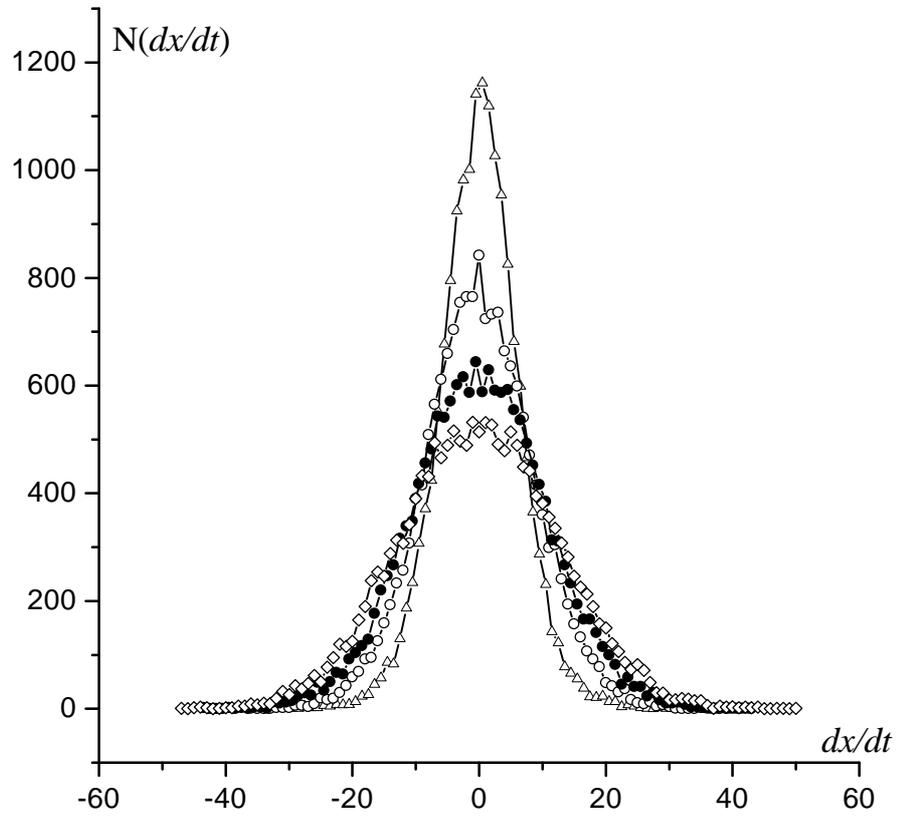
\end{document}